\documentclass{ws-ijmpd}
\usepackage[super,compress]{cite}

\usepackage{graphicx,color}

\newcommand{\be}[1]{\begin{equation}\label{#1}}
\newcommand{\ee}{\end{equation}}
\newcommand{\ba}[1]{\begin{eqnarray}\label{#1}}
\newcommand{\ea}{\end{eqnarray}}
\newcommand{\rf}[1]{(\ref{#1})}
\newcommand{\nn}{\nonumber}

\begin{document}

\markboth{Maxim Eingorn} {Cosmological Law of Universal Gravitation}

\catchline{}{}{}{}{}

\title{COSMOLOGICAL LAW OF UNIVERSAL GRAVITATION}

\author{MAXIM EINGORN}

\address{CREST and NASA Research Centers, North Carolina Central University\\ Fayetteville st. 1801,
Durham, North Carolina 27707, U.S.A.\\
maxim.eingorn@gmail.com}

\maketitle


\begin{abstract}
Without exceeding the limits of the conventional $\Lambda$CDM paradigm, we argue for Yukawa law of interparticle interaction as the law of gravitation in the real
expanding inhomogeneous Universe. It covers the whole space and comes up to take place of Newtonian gravity, which is restricted exclusively to sub-horizon
distances. The large-scale screening of gravitational interaction between every two nonrelativistic massive particles is ensured by the homogeneous cosmological
background (specifically, by the nonzero average rest mass density of nonrelativistic matter). We take advantage of the uniform matter distribution case (i.e. the
homogeneous Universe limit) to demonstrate superiority of Yukawa gravity. Attention is also devoted to the concrete particular case of inhomogeneity.
\end{abstract}

\


\ccode{PACS numbers: 98.80.Jk; 95.30.Sf; 95.35.+d; 95.36.+x; 98.65.Dx.}


\


What is the law of universal gravitation? Understanding by this law a certain well-defined formula describing the gravitational interaction between every two
massive point-like particles, let us try to answer the raised tricky question and briefly discuss the applicability bounds and a particular virtue of our
foreseeable reply. We begin with equations of motion which gravitationally interacting particles obey in the real globally expanding Universe, continue by a
decisive argument strongly corroborating the findings and conclude by an illustrative example of the nonuniform mass distribution.

\section*{Cosmological dynamics: Newtonian vs. Yukawa gravity}

It is common knowledge that if strong spacetime distortions in the vicinity of black holes or neutron stars are not at the center of attention and we restrict
ourselves to weak gravitational fields \cite{Adamek2}, then in the case of the flat Minkowski background the superposition principle holds true since there are no
cross terms in linearized Einstein equations \cite{Landau}, and the desired answer for nonrelativistic matter sounds ordinarily: Newtonian law of gravitation. As
we know since schooldays, according to this famous physical law, the gravitational potential produced by a particle of mass $m_0$ situated at the point with
radius-vector ${\bf R}_0$ has the form
\be{1} \varphi_N=-\frac{Gm_0}{|{\bf R}-{\bf R}_0|}\, ,\ee
where $G$ is the gravitational constant. Any other particle of mass $m$ at the observation point with radius-vector ${\bf R}$ experiences the action of the
corresponding force
\be{2} -m\frac{\partial\varphi_N}{\partial{{\bf R}}}=-Gm_0m\frac{{\bf R}-{\bf R}_0}{|{\bf R}-{\bf R}_0|^3}\, .\ee
Consequently, equations of motion for a finite system of particles can be written as
\be{3} \ddot {\bf R}_j=-G\sum_{i\neq j}m_i\frac{{\bf R}_j-{\bf R}_i}{\left|{\bf R}_j-{\bf R}_i\right|^3}\, ,\ee
where dots denote derivatives with respect to time $t$. The left-hand side of Eq.~\rf{3} represents the acceleration of the $j$-th particle (of mass $m_j$, with
radius-vector ${\bf R}_j$).

A manifest formidable challenge to Newtonian gravity lies in the following. First of all, in concordance with modern cosmology, Minkowski background bears no
relation to the real world and gives way to Friedmann-Lema\^{\i}tre-Robertson-Walker geometry. Besides, the number of particles (inhomogeneities in the form of
separate galaxies, their groups and clusters) in the whole cosmological system can be infinite. The global expansion of the Universe is taken into account in
computer simulation codes \cite{Springel,Dolag} for sub-horizon scales by adding an extra term in the left-hand side of Eq.~\rf{3}:
\be{4} \ddot {\bf R}_j -\frac{\ddot a}{a} {\bf R}_j=-G\sum_{i\neq j}m_i\frac{{\bf R}_j-{\bf R}_i}{\left|{\bf R}_j-{\bf R}_i\right|^3}\, , \ee
where the scale factor $a(t)$ satisfies the background Friedmann equations. In the framework of the conventional $\Lambda$CDM paradigm they may be written as
follows:
%
%
%
\be{6} \left(\frac{\dot a}{a}\right)^2=\frac{8\pi G\overline\rho}{3a^3} + \frac{\Lambda c^2}{3}\, ,\quad\quad \frac{\ddot a}{a}=-\frac{4\pi G\overline\rho}{3a^3}
+\frac{\Lambda c^2}{3}\, .\ee
Here $c$ is the speed of light while $\overline\rho$ and $\Lambda$ stand for the (constant) average rest mass density of nonrelativistic matter in comoving
coordinates and the cosmological constant, respectively. The radiation contribution has been totally disregarded. Besides, it is important to stress that each
radius-vector in Eq.~\rf{4} as well as throughout our narration is the physical (non-comoving) one.

Equations of motion similar to \rf{4} have been also analyzed in the papers \cite{EKZ2,Ellis}. The above-mentioned additional term (i.e. the second one in the
left-hand side) does not concern the law of interparticle gravitational interaction and describes the acceleration acquired by particles due to the global
cosmological expansion. If a particle is so far from its closest neighbors that their fields are negligible at its location, then such a particle obeys the
equation of motion
\be{7} \ddot {\bf R} -\frac{\ddot a}{a} {\bf R}=0\, ,\ee
asymptotically approaching the Hubble flow $\dot {\bf R}=H{\bf R}$ \cite{EZcosm1}, where $H\equiv\dot a/a$ stands for the Hubble parameter. Substitution of the
second Friedmann equation \rf{6} into \rf{7} brings to the acceleration
\be{8} \ddot {\bf R} =\left(-\frac{4\pi G\overline\rho}{3a^3} + \frac{\Lambda c^2}{3}\right) {\bf R}\, .\ee

In addition to the background geometry issue, classical (nonrelativistic) Newtonian gravity becomes inappropriate for large enough separation distances comparable
to the horizon scale where non-negligible relativistic effects (in particular, the causality issues) come into play \cite{Adamek2}. Therefore, the right-hand side
of Eqs.~\rf{3}, \rf{4} requires modification as well. Such an indispensable modification has been carried out in the recent paper \cite{Eingorn}:
\be{9} \ddot {\bf R}_j -\frac{\ddot a}{a} {\bf R}_j=-\frac{\partial \varphi_Y}{\partial {\bf R}_j}\, , \ee
where up to an additive constant
\be{10} \varphi_Y=-G\sum\limits_{i\neq j}\frac{m_i}{|{\bf R}_j-{\bf R}_i|}\exp\left(-\frac{|{\bf R}_j-{\bf R}_i|}{\lambda}\right)\, .\ee
Here we have disregarded velocities of particles as a field source \cite{Chisari}. Such a simplification is also substantiated by corresponding numerical
estimates \cite{Eingorn}. In particular, one can easily demonstrate that the ratio of the omitted velocity-dependent contributions to the dominant
velocity-independent ones (standing in the right-hand side of Eq.~\rf{9}) is of the order of the ratio of the particle peculiar velocities to the speed of light
during the entire matter-dominated and $\Lambda$-dominated stages of the Universe evolution. And this latter ratio is really very small in the analyzed case of
nonrelativistic peculiar motion.

The interaction range $\lambda$ is defined as follows:
\be{11} \lambda=\left(\frac{c^2a^3}{12\pi G\overline\rho}\right)^{1/2}\, .\ee

The prevalent weak gravitational field limit represents the only approximation, which the paper \cite{Eingorn} relies on. Without any additional assumptions,
linearized Einstein equations are solved there exactly, and the gravitational field produced by inhomogeneously distributed gravitating masses is explicitly
determined. The derived solutions including the potential \rf{10} are valid for arbitrary (sub-horizon and super-horizon) scales. Consequently, they remove
restrictions imposed on distances in modern cosmological $N$-body problems and enable running new series of high-precision simulations \cite{BrilEin}. The volume
of space covered by these simulations would be limited only by such technicalities as the computer power, but not by the underlying theory itself. This fact
represents an indubitable advantage over Newtonian equations of motion \rf{4}, which are appropriate solely for sufficiently small volumes.

Now the highway is open to us: returning to our cardinal initial question, we can asseverate that the cosmological law of universal gravitation is Yukawa law.
Really, in full accord with the expression \rf{10}, each mass produces Yukawa potential with the same finite time-dependent range $\lambda\sim a^{3/2}$ \rf{11}.

\section*{Decisive argument}

Without casting doubt on the furnished strong mathematical evidence \cite{Eingorn}, let us try to find some independent theoretical test which would corroborate
the daring idea of Yukawa gravitational interaction. Fortunately, such a crucial test does exist. Let us address the limiting case of the homogeneous mass
distribution. Then it is expected that each particle participates in the Hubble flow and, hence, obeys Eq.~\rf{7} with zero right-hand side. The question of the
proposed simple test sounds: do the right-hand sides of Eqs.~\rf{4} or \rf{9} really give zero in the investigated limiting case?

Let us start with Eq.~\rf{4} (i.e. Newtonian cosmological approximation) and consider the surface of a sphere of radius $R$. This sphere is drawn in the space
uniformly filled with matter. Then the total Newtonian force induced by the outer space with respect to the outlined surface is zero (when integrating over a
sequence of concentric shells) while the inner space generates the nonzero acceleration $-4\pi G\overline\rho {\bf R}_j/\left(3a^3\right)$. Substituting it
instead of the right-hand side of Eq.~\rf{4} and omitting the irrelevant subscript $j$, we obtain
\be{12} \ddot {\bf R} -\frac{\ddot a}{a} {\bf R}=-\frac{4\pi G\overline\rho}{3a^3}{\bf R}\, ,\ee
or, after substitution of the second Friedmann equation \rf{6},
\be{13} \ddot {\bf R}=\left(-\frac{8\pi G\overline\rho}{3a^3} + \frac{\Lambda c^2}{3}\right){\bf R}\, .\ee
This means that the matter contribution is groundlessly doubled in comparison with the correct acceleration \rf{8}.

The subtlety of this result consists in the fact that it depends on the order of integration. In this connection, a different specific order \cite{Peebles} may
assure zero in the right-hand side of Eq.~\rf{12}. Nevertheless, if one insists on the physically motivated idea that the well-defined total force should not
depend on the way of summing up forces induced by individual gravitating masses, then the discussed drawback of Newtonian cosmological approximation becomes
evident.

On the contrary, if Eq.~\rf{9} (i.e. Yukawa gravity) is at the center of attention, then combined contributions from the inner and outer spatial regions reduce to
zero irrespectively of the integration sequence. Indeed, in order to demonstrate this, we can use the formula for the radial acceleration of a test body within a
uniformly filled spherical shell of inner and outer radii $R_1$ and $R_2$, respectively, concentrating solely on the Yukawa part and, hence, eliminating Newtonian
trace \cite{shell}:
\ba{14} -\frac{\partial\varphi_Y}{\partial R}=&-&\frac{4\pi
G\overline\rho\lambda^3}{a^3R^2}\left[h\left(\frac{R}{\lambda}\right)\left(1+\frac{R_2}{\lambda}\right)\exp\left(-\frac{R_2}{\lambda}\right)\right.\nn\\
&-&\left. h\left(\frac{R_1}{\lambda}\right)\left(1+\frac{R}{\lambda}\right)\exp\left(-\frac{R}{\lambda}\right)\right]\, ,\ea
where
\be{15} h\left(\chi\right)\equiv \chi\cosh\left(\chi\right)-\sinh\left(\chi\right)\, .\ee
The homogeneous Universe corresponds to the simultaneous limits $R_1\rightarrow0$ and $R_2\rightarrow+\infty$ reducing \rf{14} to zero. Thus, the desired equation
of motion \rf{7} is reinstated, strongly corroborating superiority of Yukawa gravitation law.

\section*{Illustrative example}

It is also noteworthy that inside a solitary sphere of radius $R_1$, being completely empty with the exception of its central point where the mass $M=4\pi
\overline\rho R_1^3/\left(3a^3\right)$ is resting (Einstein-Straus/Swiss-cheese models \cite{ES,EZcosm2,EBV,SC}), the external homogeneous spatial region leads to
the nonzero radial acceleration
\be{16} -\frac{\partial\varphi_Y^{\mathrm{(in)}}}{\partial R}=\frac{4\pi
G\overline\rho\lambda^3}{a^3R^2}h\left(\frac{R}{\lambda}\right)\left(1+\frac{R_1}{\lambda}\right)\exp\left(-\frac{R_1}{\lambda}\right)\, ,\ee
which for $R_1\ll\lambda$ (and, hence, $R\ll\lambda$ since $R<R_1$ for the internal space) takes the form $4\pi G\overline\rho R/\left(3a^3\right)$. Therefore, it
compensates exactly the matter part nestling in Eq.~\rf{9} within the term $-(\ddot a/a)R$, resulting in the equation of motion
\be{17} \ddot {\bf R} =-\frac{GM}{R^2}\frac{{\bf R}}{R}+\frac{\Lambda c^2}{3} {\bf R} \ee
and thereby confirming that the global Universe expansion affects the motion of a test body inside the investigated sphere through the instrumentality of the
cosmological constant $\Lambda$ only, in solid agreement with the famous Schwarzschild-de Sitter metric \cite{ES}. However, if one resorts to Newtonian equations
of motion \rf{4} naively disregarding the external region contribution, then
\be{17a} \ddot {\bf R} =-\frac{GM}{R^2}\frac{{\bf R}}{R}+\left(-\frac{4\pi G\overline\rho}{3a^3} + \frac{\Lambda c^2}{3}\right) {\bf R}\, , \ee
where the groundless additional term $-\left[4\pi G\overline\rho/\left(3a^3\right)\right]{\bf R}$ arises in the right-hand side due to lack of compensation
mechanism.


For the sake of completeness, let us consider the general case of an arbitrary ratio $R_1/\lambda$. Then the equation of motion \rf{17} should be rewritten for
the internal spatial region ($R<R_1$) as follows:
\be{18} \ddot {\bf R} =-\frac{\partial\varphi_Y^{\mathrm{(M)}}}{\partial R}\frac{{\bf R}}{R}+\left(-\frac{4\pi G\overline\rho}{3a^3} + \frac{\Lambda
c^2}{3}\right) {\bf R}-\frac{\partial\varphi_Y^{\mathrm{(in)}}}{\partial R}\frac{{\bf R}}{R}\, ,\ee
where the last term in the right-hand side is determined by \rf{16} while the first one describes the contribution of the central mass:
\be{19} -\frac{\partial\varphi_Y^{\mathrm{(M)}}}{\partial R}=-\frac{GM}{R^2}\left(1+\frac{R}{\lambda}\right)\exp\left(-\frac{R}{\lambda}\right)\, .\ee
At the same time, in the region $R>R_1$ we have
\be{20} \ddot {\bf R} =-\frac{\partial\varphi_Y^{\mathrm{(M)}}}{\partial R}\frac{{\bf R}}{R}+\left(-\frac{4\pi G\overline\rho}{3a^3} + \frac{\Lambda
c^2}{3}\right) {\bf R}-\frac{\partial\varphi_Y^{\mathrm{(out)}}}{\partial R}\frac{{\bf R}}{R}\, ,\ee
where the last term in the right-hand side is now determined by \rf{14} in the limit $R_2\rightarrow+\infty$:
\be{21} -\frac{\partial\varphi_Y^{\mathrm{(out)}}}{\partial R}=\frac{4\pi
G\overline\rho\lambda^3}{a^3R^2}h\left(\frac{R_1}{\lambda}\right)\left(1+\frac{R}{\lambda}\right)\exp\left(-\frac{R}{\lambda}\right)\, .\ee
Evidently, the expressions \rf{18} and \rf{20} are continuous on the surface of the sphere under consideration (that is at the distance $R=R_1$ from its center).

One can easily receive evidence that the formula \rf{18} is closely approximated by the formula \rf{17} even if the inequality $\left(R_1/\lambda\right)\ll1$ does
not hold true. Returning to the equation of motion \rf{17}, we also see that it actually lays the foundation for the three-dimensional method employed, e.g., in
the recent paper \cite{Banik} for investigating dynamics of our Local Group of galaxies. The authors exclude the part $-4\pi G\overline\rho/\left(3a^3\right)$
from $\ddot a/a$ while keeping the part $\Lambda c^2/3$ untouched, and appeal to a finite system of gravitationally interacting particles in the empty Universe in
presence of the cosmological constant. Now a different interpretation is available: the matter contribution in $\ddot a/a$ is exactly compensated by the
corresponding total Yukawa contribution from the external space treated as homogeneous beyond the analyzed system of particles.




\section*{Conclusion}

We summarize by reasserting that Yukawa potential, which is inherent in elementary particle and plasma physics, surprisingly governs universal gravitation as
well. The cosmological screening length $\lambda$ is determined by the average rest mass density of nonrelativistic matter by means of the definition \rf{11} and
amounts to $3.7$~Gpc at present, giving estimate of the homogeneity scale/upper limit of the cosmic structure dimension/bound to a spatial domain of probable
structure development \cite{Eingorn,Ruslan,Kiefer}. When viewed at a scale greater than $\lambda$, the Universe is homogeneous and isotropic, without a trace of
intensive galactic clustering, in complete agreement with the basic cosmological principle and confirmative cosmic microwave background and other observational
data. Armed with the achieved results, which may be utterly important in the light of the precision cosmology era and future surveys such as Euclid, we pretend to
accept various physical challenges.

\section*{Acknowledgements}

This work was partially supported by NSF CREST award HRD-1345219 and NASA grant NNX09AV07A.


\end{document}